\documentstyle[aasms4,12pt,flushrt]{article}

\def\ifm#1{\relax\ifmmode#1\else$\mathsurround=0pt #1$\fi}

\begin{document}

\title{The Unusual Spectral Energy Distribution of a Galaxy Previously Reported
to be at Redshift 6.68}

\author{Hsiao-Wen Chen\altaffilmark{1}, Kenneth M. Lanzetta\altaffilmark{2}, 
Sebastian Pascarelle\altaffilmark{2}, \& Noriaki Yahata\altaffilmark{2}}

{\small \noindent $^1$Observatories of the Carnegie Institution of Washington,
813 Santa Barbara, Pasadena, CA 91101, USA}

{\small \noindent $^2$Department of Physics and Astronomy, State University of
New York at Stony Brook, Stony Brook, NY 11794--3800, USA}

{\bf \noindent Observations of distant galaxies are important both for 
understanding how galaxies form and for probing the physical conditions of the 
universe at the earliest epochs.  It is, however, extremely difficult to 
identify galaxies at redshift $z>5$, because these galaxies are faint and 
exhibit few spectral features.  In a previous work, we presented observations 
that supported the identification of a galaxy at redshift $z = 6.68$ in a deep 
STIS field$^1$.  Here we present new ground-based photometry of the galaxy.  We
find that the galaxy exhibits moderate detections of flux in the optical $B$ 
and $V$ images that are inconsistent with the expected absence of flux at 
wavelength shortward of the redshifted Lyman-$\alpha$ emission line of a galaxy
at redshift $z>5$.  In addition, the new broad-band imaging data not only show 
flux measurements of this galaxy that are incompatible with the previous STIS 
measurement, but also suggest a peculiar spectral energy distribution that 
cannot be fit with any galaxy spectral template at any redshift.  We therefore 
conclude that the redshift identification of this galaxy remains undetermined.}

  The galaxy (designated galaxy A by Chen, Lanzetta, \& Pascarelle$^1$) was 
previously identified in deep observations obtained by the Hubble Space 
Telescope using the Space Telescope Imaging Spectrograph (STIS) in 1997
December.  It has a STIS clear magnitude of $AB({\rm clear}) = 27.67 \pm 0.09$.
The spectrum of galaxy A was extracted using a new spectrum extraction 
technique that employed optimal weights to spatially deblend and spectrally 
deconvolve the slitless spectra.  The spectrum was characterized by a 
discontinuity at wavelength $\lambda \approx 9300$ \AA, which we interpreted 
as the Lyman-$\alpha$ decrement (produced by intervening Hydrogen absorption), 
and by an emission line at wavelength $\lambda \approx 9334$, which we 
interpreted as Lyman-$\alpha$.  But because of the low signal-to-noise ratios 
of the identified spectral features, the redshift identification of galaxy A 
was not indisputable.

  It is extremely difficult to obtain a confirming spectrum even with a 10 m 
class telescope on the ground because, at near-infrared wavelengths, background
sky light is the dominant source of noise and confusing sky emission lines make
identifications of spectral line features ambiguous.  To strengthen the 
identifications of the faint galaxies observed in the STIS field, we acquired 
deep optical $B$, $V$, and $R$ images of the STIS field with the WIYN telescope
at the Kitt Peak National Observatory using the Mini Mosaic camera.  The 
primary objective was to obtain sensitive photometry over a wide 
passband in order to provide strong constraints in the spectral energy 
distributions of these galaxies.  In particular, a complete absence of flux at 
optical wavelengths shortward of the redshifted Lyman-$\alpha$ emission line 
due to intervening Hydrogen absorption would unambiguously confirm the previous
redshift identification of galaxy A.  The observations were carried out in a 
series of exposures of between 900 and $1,200$ s each in 2000 January.  The 
total exposure times of the $B$, $V$, and $R$ images were 2.3, 2.0, and 3.5 h, 
which reached 5 $\sigma$ point source limiting $AB$ magnitudes of $\approx$ 
26.3, 26.0, and 26.7, respectively.  The individual exposures were reduced 
using standard pipeline techniques and were registered to a common origin and 
filtered for cosmic rays.  

  To measure the energy fluxes of galaxies in the STIS field, we adopted a 
quasi-optimal photometry technique that fits model spatial profiles of detected
objects to the individual ground-based images$^2$.  The model spatial profiles 
were obtained by convolving smooth models of the intrinsic profiles of detected
objects (determined based on the direct image previously obtained with STIS 
using a non-negative least-squares image reconstruction method$^3$) with 
appropriate point spread functions of individual images (approximated using a 
Gaussian profile).  The new technique provides higher signal-to-noise ratio and
more accurate flux uncertainty measurements than standard techniques, because 
it properly accounts for the variation of the point spread functions of 
individual frames (the full width at half maximum of which was measured to vary
between 0.5 and 1.0 arcsec) and because it accounts for uncertainty 
correlations between nearby, overlapping neighbors.

  We present the new ground-based $B$, $V$, and $R$ images of galaxy A in 
Figure 1 together with the direct image obtained in the STIS clear filter.
Contrary to expectations based on the previous analysis, galaxy A exhibits 
moderate detections of flux in all three ground-based images.  Specifically, we
estimate that galaxy A has $B$, $V$, and $R$ magnitudes of $AB(B) = 26.7 \pm
0.2$, $AB(V) = 26.8\pm0.4$, $AB(R) = 27.3\pm0.4$, respectively.  For a galaxy 
at redshift $z = 6.68$, both the $B$ and $V$ filters cover a wavelength range 
below the redshifted Lyman-$\alpha$ emission line.  Therefore, the presence of 
optical flux in the $B$ and $V$ images indicates that galaxy A is very unlikely
to be at redshift $z = 6.68$.  Nevertheless, it is puzzling that the 
ground-based photometry do not seem to agree with the previous STIS photometry.
Specifically, we find that for a STIS clear magnitude of $AB({\rm clear}) = 
27.67 \pm 0.09$, galaxy A is at least a factor of two too bright in the $B$ and
$V$ filters when compared with known galaxy spectral energy distributions of 
any given redshift.  

  Figure 2 shows the spectral energy distribution of galaxy A established based
on the broad-band photometry in the $B$, $V$, $R$ (open circles),
and STIS clear filters (open square) in comparison to the extracted 
one-dimensional STIS spectrum of galaxy A cast into 350 \AA\ bins (open 
triangles).  The spectral coverage of each filter is indicated by horizontal
bars.  It further shows that galaxy A has an extremely blue color, $AB(B)-AB(R)
= -0.6 \pm 0.4$.  The only alternative for the new photometry
to be in line with the previous measurement would be that a dominant portion of
the flux observed in the STIS clear filter occurred within the spectral range 
covered by the $B$, $V$, and $R$ filters (i.e.\ between 3500 and 9000 \AA) and 
a negligible amount of flux at wavelengths below or beyond this range.  The 
closest resemblance would be the spectral energy distribution of an F type 
star, but the apparent magnitude of this object would imply a distance of 
$\approx 450$ kpc from the Sun, many times larger than the size of the Milky 
Way.  We therefore conclude that this cannot be the case.

  Because the STIS direct image (covering a wavelength range that spans from 
2000 \AA\ to 10000 \AA) is a factor of $\approx$ 10 times more sensitive than 
the ground-based images and because STIS photometric calibration is accurate to
within 5\% (according to the STIS Instrument Handbook), it ought to provide a 
strong, reliable constraint in the integrated optical flux of galaxy A.  To 
resolve the discrepancy between the ground-based and STIS photometry for galaxy
A, we first examined the photometric calibration of the ground-based images.  
First, we measured photometry in the $B$, $V$, and $R$ images for all the 1067 
objects in the Hubble Deep Field (HDF) region presented by Fern\'andez-Soto, 
Lanzetta, \& Yahil$^4$.  Next, we converted the ground-based $B$, $V$, and $R$ 
photometry to the space-based F450W and F606W photometry using the 
transformation presented in Holtzman et al.$^5$  Finally, we compared our 
photometry of these objects with the ones presented in 
Fern\'andez-Soto, Lanzetta, \& Yahil$^4$.  We found that these measurements 
completely agreed with each other.  We therefore concluded that the photometric
calibration of the ground-based images was consistent with the space-based 
photometry and that the discrepant photometry was due to some
other factor.  Because the two sets of observations were taken two years apart,
the discrepant photometry could suggest that galaxy A is 
variable.  In particular, the observed blue $V-R$ color of galaxy A agrees with
that of the supernova 1995K identified at $z=0.479$$^6$.  But we cannot provide
further support to the hypothesis.

  Regardless of the peculiarity of the spectral energy distribution of galaxy 
A, the conflicting results from two separate analyses also suggest that the 
previous analysis of the STIS observations might be in error.  We therefore 
examined carefully the previous analysis, particularly the spectral extraction 
technique.

  The primary arguments that were presented by Chen, Lanzetta, \& 
Pascarelle$^1$ to support the confidence of the redshift identification were 
(1) consistent flux measurements from the direct image ($AB({\rm clear}) = 
27.67 \pm 0.09$) and the dispersed image ($AB({\rm clear}) = 27.72 \pm 0.29$), 
(2) a robust detection of the emission line at $\lambda \approx 9334$, and (3) 
a statistically significant detection of the flux discontinuity at $\lambda 
\approx 9300$ \AA\ that coincided with the emission line.  We confirmed the
detection of the emission line and the flux discontinuity using traditional
spectral extraction techniques, but we were unable to confirm the amplitude of
the flux discontinuity.  The amplitude of the flux discontinuity could be 
overestimated in the previous analysis due to small systematic errors in the 
sky determination of the dispersed image, but we do not find evidence for that.

  The results of our analysis show that galaxy A exhibits peculiar spectral 
properties that cannot be explained by either galaxies or stars.  We cannot 
resolve the incompatible photometry of the new ground-based 
images and the previous STIS images.  Therefore, we conclude that the redshift 
identification of galaxy A remains undetermined.

\vspace{0.1in}

\hrule

\begin{small}

\noindent Received {\underline{\makebox[1in]{}}.}

\noindent 1. Chen, H.-W., Lanzetta, K. M., Pascarelle, S. A Spectroscopically 
Identified Galaxy of Probable Redshift $z=6.68$. Nature {\bf 398}, 586--588 
(1999).

\noindent 2.  Yahata, N., Lanzetta, K. M., Chen, H.-W., Fern\'andez-Soto, A., 
Pascarelle, S. M., Puetter, R. C., \& Yahil, A. Photometry and Photometric 
Redshifts of Faint Galaxies in the Hubble Deep Field South NICMOS Field. 
Astrophys.\ J., {\bf 538}, 493--504 (2000).

\noindent 3. Puetter, R. C. \& Yahil, A. The Pixon Method of Image 
Reconstruction in ``The Astronomical Data Analysis Software and Systems (ADASS)
VIII Conference,'' eds. D. M. Mehringer, R. L. Plante, \& D. A. Roberts (San 
Francisco: ASP), 307--316

\noindent 4. Fern\'andez-Soto, A., Lanzetta, K. M., \& Yahil, A. A new catalog
of photometric redshifts in the Hubble Deep Field. Astrophys.\ J., {\bf 513},
34--50 (1999).

\noindent 5. Holtzman, J. A., Burrows, C. J., Casertano, S., Hester, J. J., 
Trauger, J. T., Watson, A. M., \& Worthey, G. The photometric Performance and 
Calibration of WFPC2. PASP, {\bf 107}, 1065--1093 (1995).

\noindent 6. Schmidt, B. P. et al.\ The High-z Supernova Search: Measuring
Cosmic Deceleration and Global Curvature of the Universe Using Type Ia 
Supernovae. Astrophys.\ J., {\bf 507}, 46--63 (1998)

%\noindent 6. Lanzetta, K. M., Yahil, A., \& Fern\'andez-Soto, A. Star-forming
%galaxies at very high redshifts. Nature {\bf 381}, 759--763 (1996).

%\noindent 7. Guzm\`an, R., Gallego, J., Koo, D. C., Phillips, A. C., Lowenthal,
%J. D., Faber, S. M., Illingworth, G. D., \& Nicole, P. V. The nature of compact
%galaxies in the Hubble Deep Field. II. Spectroscopic properties and 
%implications for the evolution of the star formation rate density of the 
%universe. Astrophys.\ J. {\bf 489}, 559--572 (1997).
\end{small}

\noindent CORRESPONDENCE should be addressed to H.-W.C. (email:
hchen@\-ociw.\-edu).

\newpage

\figcaption{Portions of the ground-based $B$, $V$, and $R$ images, smoothed by 
the size of the point spread functions, and the corresponding STIS image 
centered at galaxy A.  Each panel is 14 arcsec on a side.  North is to the left
and East is down.  The $AB$ magnitudes of galaxy A are $AB(B) = 26.7\pm0.2$, 
$AB(V) = 26.8\pm0.4$, $AB(R) = 27.3\pm0.4$, and $AB({\rm clear}) = 27.67 \pm 
0.09$.}

\figcaption{The observed spectral energy distribution of galaxy A.  The open
circles represent the broad-band photometry in the WIYN $B$, $V$,
and $R$ images.  The open square represents the photometric measurement in the
STIS clear filter.  The open triangles represent the extracted one-dimensional 
STIS spectrum of galaxy A cast into 350 \AA\ bins.  The spectral coverage of 
each filter is indicated by horizontal bars. }

\newpage

\plotone{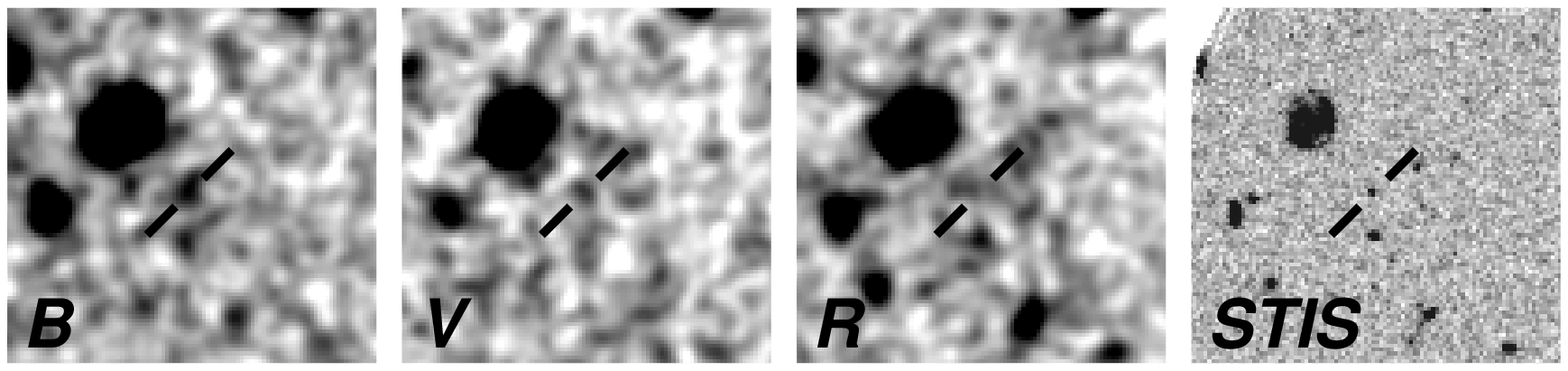}

\plotone{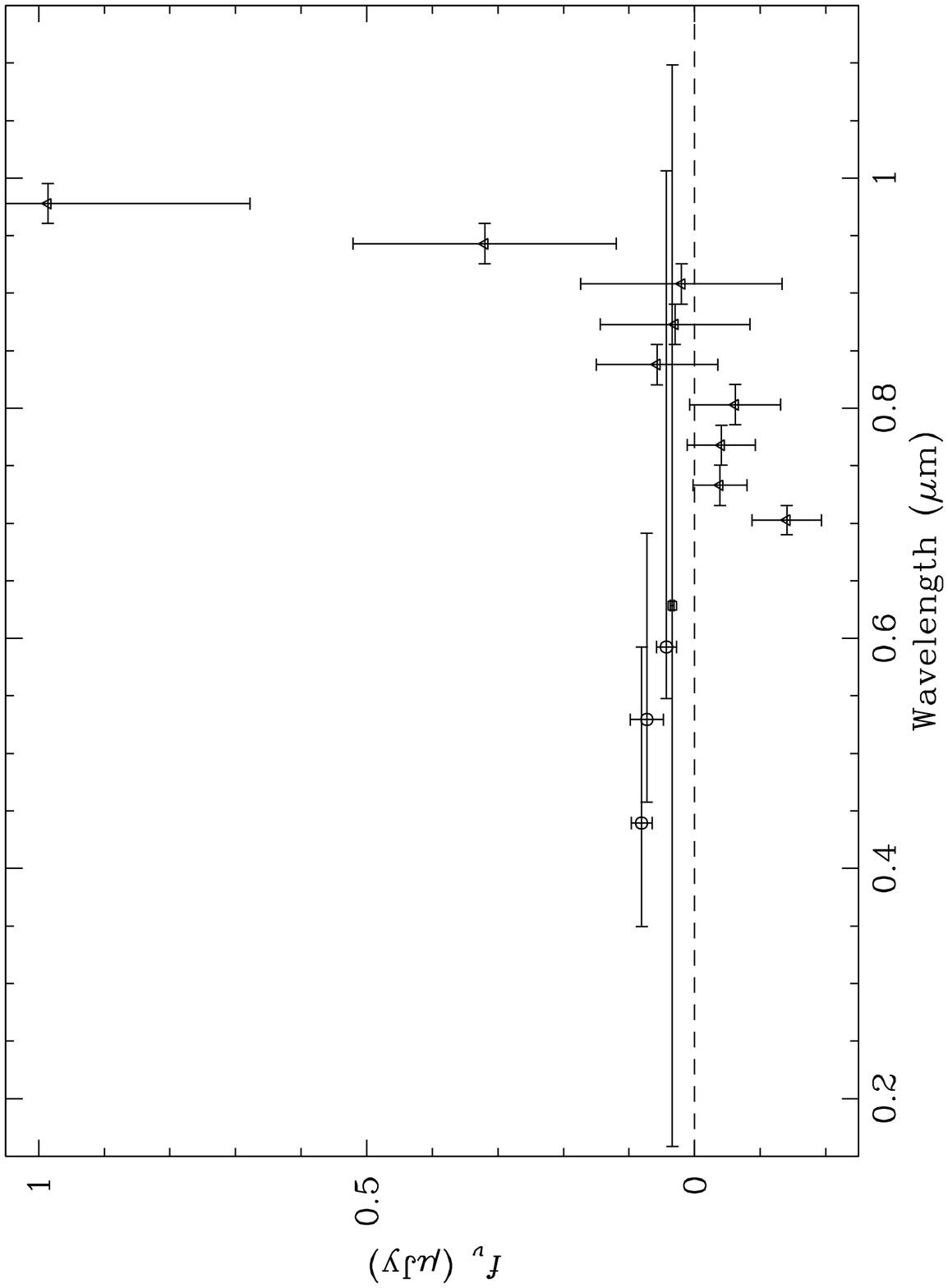}

\end{document}